# Reliable Magnetometry for Antiferromagnets and Thin Films: Correcting Substrate Artifacts in Mn$_3$Sn/MgO Systems

Katarzyna Gas[1,2,*] and Maciej Sawicki[2,3,†]

[1] *Center for Science and Innovation in Spintronics, Tohoku University, 2-1-1 Katahira, Aoba-ku, Sendai 980-8577, Japan*
[2] *Institute of Physics, Polish Academy of Sciences, Aleja Lotnikow 32/46, PL-02668 Warsaw, Poland*
[3] *Laboratory for Nanoelectronics and Spintronics, Research Institute of Electrical Communication, Tohoku University, 2-1-1 Katahira, Aoba-ku, Sendai 980-8577, Japan*

**ABSTRACT**. The rapid progress in antiferromagnetic and altermagnetic spintronics has led to increased interest in magnetic materials with vanishing net magnetization but strong spin-dependent transport properties. As thin films of such materials become central to device concepts, precise magnetic characterization is essential, for quantify intrinsic moments, and interpret transport signatures such as the anomalous Hall effect. In this work, we show that commercial MgO substrates, commonly used in epitaxial growth, often produce substantial parasitic magnetic signals that can match or exceed the response of weakly magnetic films. We identify two major components: a low-field ferromagnetic-like contribution originating from epi-ready surface, and a temperature-dependent paramagnetic background associated with dilute bulk impurities. These artifacts vary between samples and cannot be corrected using standard linear background subtraction or a measured reference substrate. To address this, we develop and put forward a compensation scheme which combines two complementary, non-destructive measurement protocols. We demonstrate up to 97% efficacy without requiring prior measurements of the bare substrate. The proposed framework enables reliable extraction of intrinsic magnetic signals and provides a general strategy for high-fidelity magnetometry in weakly magnetic thin-film systems, including emerging classes of materials such as topological phases and two-dimensional magnets. We also report the diamagnetic susceptibility of crystalline MgO, $\chi_{\mathrm{MgO}} = -4.0 \times 10^{-7}$ emu/g/Oe.

*Contact author: gas.katarzyna.a2@tohoku.ac.jp
†Contact author: mikes@ifpan.edu.pl

## I. INTRODUCTION.

Antiferromagnetic (AFM) materials are increasingly studied for spintronic devices due to fast dynamics, robustness against magnetic fields, and large magnetotransport responses despite a vanishingly small net magnetization [1,2]. Recent developments, including the emergence of altermagnets (AM) with compensated order and spin-polarized bands have accelerated research [3,4]. Consequently, reliable magnetic characterization in thin-film form is essential. The case of RuO$_2$, a candidate AM with highly debated magnetic ground state, further underscores the importance of artifact-resistant measurements methods [5].

This task is particularly demanding for systems with near-zero net magnetization, such as antiferromagnets and ultrathin films. While typical ferromagnetic layers exhibit magnetic moments corresponding to flux densities of the order of tesla, antiferromagnets – even those with weak ferromagnetism – often yield only millitesla range signals [6–9]. These must be extracted from a much larger substrate's signal, which usually dominates the measured moment. Accurate correction is thus critical to compare experiment with theory, identify switching events, domain states, extract intrinsic magnetic moments, and to ensure that correlations with transport effects, including topological ones, reflect genuine material behavior rather than substrate-induced artifacts [10].

Similar challenges have previously arisen in studies of dilute magnetic semiconductors (DMS), where claims of (room temperature) ferromagnetism, FM, were often later attributed to unintentional formation of secondary ferromagnetic phases rather than intrinsic magnetic ordering [11] or contamination [12–15]. Appendix A summarizes common sources of such spurious magnetic contributions in thin-film measurements and provides key references on identifying, avoiding, and mitigating major artifacts. These historical examples reemphasize the need for meticulous preparation, contamination control, and reliable background subtraction.

A commonly used approximation assumes that the substrate behaves as an ideal diamagnet with a linear in magnetic field, $H$, and temperature, $T$, independent magnetic response. In practice, however, real substrates often exhibit nonlinearities and sample-specific effects that invalidate this assumption. Even high-quality wafers such as Si, GaAs, or $Al_2O_3$ can show deviations from ideal diamagnetism at the $10^{-6}$ emu level or above [16–18]. Crystalline oxide substrates, including MgO, $SrTiO_3$, LSAT, and $MgAl_2O_4$, are especially problematic due to paramagnetic and ferromagnetic-like contributions [19].

In this work, we investigate the detrimental role of commercial crystalline substrates using thin films of AFM $Mn_3Sn$ grown epitaxially on MgO as a representative model system. In its bulk crystal form, $Mn_3Sn$ is a well-established noncollinear antiferromagnet with an inverse triangular spin arrangement in the kagome plane and weak ferromagnetic-like magnetization [2 m$\mu_B$ per Mn atom, or 6 m$\mu_B$/f.u. (per formula unit), where $\mu_B$ is the Bohr magneton, or 0.9 mT in SI units] due to subtle canting of Mn moments[16,17]. Orientation-dependent studies reveal strong magnetic anisotropy [20]. When the magnetic field is applied within the kagome plane ($H \parallel [2\bar{1}\bar{1}0]$ or $H \parallel [01\bar{1}0]$), the remnant magnetization $M_r$ is nearly equal to the spontaneous value, indicating robust and well-defined magnetic switching [20]. In contrast, for $H \parallel [0001]$ magnetization is linear in $H$, with no hysteresis or remanence. Temperature-dependent studies show that both $M_r$ and $H_c$ decrease gradually from 300 K and vanish near 420 K, consistent with the Néel temperature $T_N$ [20]. In bulk crystals, the anomalous Hall effect (AHE) and magnetization loops generally show similar features with matching coercive fields and squareness, suggesting a shared microscopic origin in the chiral spin structure [20]. While the AHE does not scale directly with net magnetization, this qualitative agreement makes it a useful proxy for magnetization reversal and domain dynamics. Subtle deviations between AHE and $M(H)$ have also been reported and, in certain cases, attributed to additional topological contributions – specifically, to real-space Berry curvature from chiral spin textures such as noncoplanar domains [21].

Our motivation stems from the observation that the correspondence between AHE and magnetization becomes inconsistent in thin-film systems. Some studies on $Mn_3Sn$ thin films have reported agreement between these two quantities [22–25], while others reveal striking inconsistencies [26–28]. For example, AHE loops often exhibit substantial coercivity and remanence, while the corresponding magnetization loops are nearly non-hysteretic and show anomalously large saturation values. If these differences are addressed, they are typically attributed to interfacial contributions, uncompensated surface spins, defects or to the commonly cited notion that AHE and net magnetization have different origins, without explaining the divergence in switching behavior. Therefore, understanding the relationship between AHE and magnetization in AFM thin films remains an important question for both fundamental physics and device applications.

Moreover, many experimental studies lack clear documentation on substrate correction [26,27,29–32]. When attempted, corrections are often limited to subtracting a 'diamagnetic background' [33–35] – a linear-in-field contribution fitted to high-field data. This approach assumes the ideally diamagnetic magnetic response and so neglects, without verification, the presence of sample-specific $T$- and/or $H$-dependent nonlinear magnetic components. Such unaccounted contributions can distort the extracted shape of the film's magnetization curve, as they are not eliminated by a simple linear background subtraction.

Given these inconsistencies, and drawing on lessons from DMS studies, we now systematically investigate how to isolate the intrinsic magnetic response of $Mn_3Sn$ thin films from extrinsic contributions. By using the well-understood bulk AHE-magnetization correspondence as a guiding reference, we identify deviations that arise specifically from artifacts introduced by overly simplified data reduction procedures. This makes $Mn_3Sn$ an ideal platform for validating background correction methods applicable to all weakly magnetic thin-film systems.

We show that MgO substrates contribute two distinct, sample-specific components superimposed on the intrinsic diamagnetic background: a broad paramagnetic (PM) contribution and a low-field ferromagnetic-like component (FML). These nonlinear terms can match or exceed the film's signal and vary between nominally identical substrates, precluding a universal correction based on a single reference.

We introduce two complementary, non-destructive correction methods for final film measurements without needing prior bare-substrate data:
• The Internal Reference Method (IRM), subtracting high-temperature magnetization loop – where the film contribution is negligible – to remove the FML component, and

- The Differential Isotherm Method (DIM), modeling the temperature dependent PM component.

IRM restores agreement between magnetometry and AHE coercive fields and suppresses unphysical "double-switching" artifacts. DIM enables robust PM background modeling even without remeasuring the substrate. Combined, these methods improve thin-film magnetometry by recovering physically meaningful magnetization values.

Although we focus on $Mn_3Sn$ films on MgO, these methods apply broadly to other magnetic thin films or 2D systems stabilized on commercial oxide substrates – including $SrTiO_3$, LSAT, $Al_2O_3$, and others where correction is essential. As the search for new AFM and AM materials accelerates, especially in ultrathin films, this correction framework will be crucial for accurate, reproducible magnetometry in spintronic systems, ensuring transport phenomena such as AHE, magnetoresistance, or current-induced switching can be reliably correlated with the underlying magnetic state.

## II. METHODS

$Mn_{3+x}Sn_{1-x}$ thin films with $(10\bar{1}0)$ out-of-plane orientation (*m*-plane in hexagonal notation) are grown on single-crystal MgO(110) substrates using DC/RF magnetron sputtering, following the procedure described in [36–38]. The multilayer stack, fabricated in a single vacuum cycle, consists of a W (2 nm)/Ta (3 nm) seed bilayer, the $Mn_{3+x}Sn_{1-x}$ magnetic layer, and a MgO (1.3 nm)/Ru (1 nm) capping bilayer; (thicknesses in parentheses). All layers except MgO are deposited at 400 °C; the MgO cap is deposited at room temperature. After deposition the samples are annealed in vacuum at 600°C for an hour.

Two films with different composition and thickness are studied: Sample A - $Mn_{3.02}Sn_{0.98}$ thickness 40 nm, Néel temperature $T_N$ = 385 K [25], and Sample B - $Mn_{3.09}Sn_{0.91}$, thickness 80 nm, $T_N$ = 380 K. The chemical composition is determined by inductively coupled plasma (ICP) spectroscopy on co-deposited reference samples. The Néel temperatures are determined from temperature-dependent magnetization measurements following ref. [25]. X-ray diffraction measurements (not shown) confirmed the $(10\bar{1}0)$ orientation of the $Mn_{3+x}Sn_{1-x}$ films and the absence of secondary phases within the detection limit. Single-crystal MgO substrates are obtained from a commercial vendor.

Electrical transport measurements are performed using the van der Pauw method with a manual prober equipped with a perpendicular magnetic field up to 10 kOe ($H \perp$ film plane) from Toei Scientific Industrial Co. DC excitation is used throughout the experiments.

Magnetic measurements are performed using a Quantum Design MPMS XL superconducting quantum interference device (SQUID) magnetometer equipped with the reciprocating sample option (RSO), which provides enhanced signal-to-noise ratio. The measurements are carried out in the temperature range of 2 – 400 K using standard cut silicon sticks (approximately 8″ in length) as sample holders [14]. Typically, 5 × 5 $mm^2$ specimens are fixed at the center of the stick using strongly diluted GE varnish [39].

The use of Si sticks, routinely employed by the authors in their most sensitive magnetometry studies [40–42], completely eliminates position-dependent magnetic signals commonly observed when plastic drinking straws are used as sample holders in SQUID magnetometers [43,44]. This issue becomes particularly critical for measurements above 350 K due to the thermoplastic nature of the straws, which compromises mechanical stability and reproducibility. Furthermore, based on the authors' experience, the use of plastic straws-especially "straight from the box," i.e., without prior selection or testing-is highly inadvisable and should be avoided in precision, high-sensitivity magnetometry. In this context, both silicon sticks and quartz paddle holders (as used in QD VSM-SQUID systems) provide not only excellent mechanical and thermal stability up to 400 K, but also enable highly reproducible sample positioning [45,46]. These factors are essential in experiments where the final magnetic signal corresponds to a small difference between two large background contributions.

All measurements have been carefully planned and conducted following established protocols for high-precision and high-sensitivity integral SQUID magnetometry to ensure accurate determination of weak magnetic signals, minimize experimental artifacts, and guarantee full traceability of the measured magnetic response [14].

## III. RESULTS AND DISCUSSION

### A. Linear correction in magnetic data reduction

As a starting point, we examine whether a commonly used linear background correction in the form of $\alpha \cdot H$ is truly sufficient to reconcile magnetometry and transport data in $Mn_3Sn$ thin films grown on MgO. This approach is widely used to remove substrate contributions and recover the intrinsic magnetic behavior of the film, but it relies on two key assumptions: ideal diamagnetism of the MgO substrate and complete saturation of the magnetization of $Mn_3Sn$ in strong fields. To assess its validity, we compare representative magnetometry and transport results obtained under identical conditions. Specifically, Fig. 1 illustrates the scale of discrepancies by juxtaposing the magnetic field dependence of anomalous Hall resistance, $R_{AH}(H)$ (top panels), and magnetization, $m(H)$ (bottom panels), measured at 300 K for two $Mn_3Sn$ samples studied here: Sample A (40 nm, $Mn_{3.02}Sn_{0.98}$) and Sample B (80 nm, $Mn_{3.09}Sn_{0.91}$), both grown on MgO substrates. The Hall data are shown after basic symmetrization only, while the magnetometry results represent the non-linear contributions, $m_{NL}(H)$, obtained by subtracting a linear background term, $\alpha \cdot H$, from the raw data, with $\alpha$ adjusted to achieve apparent saturation at high fields.

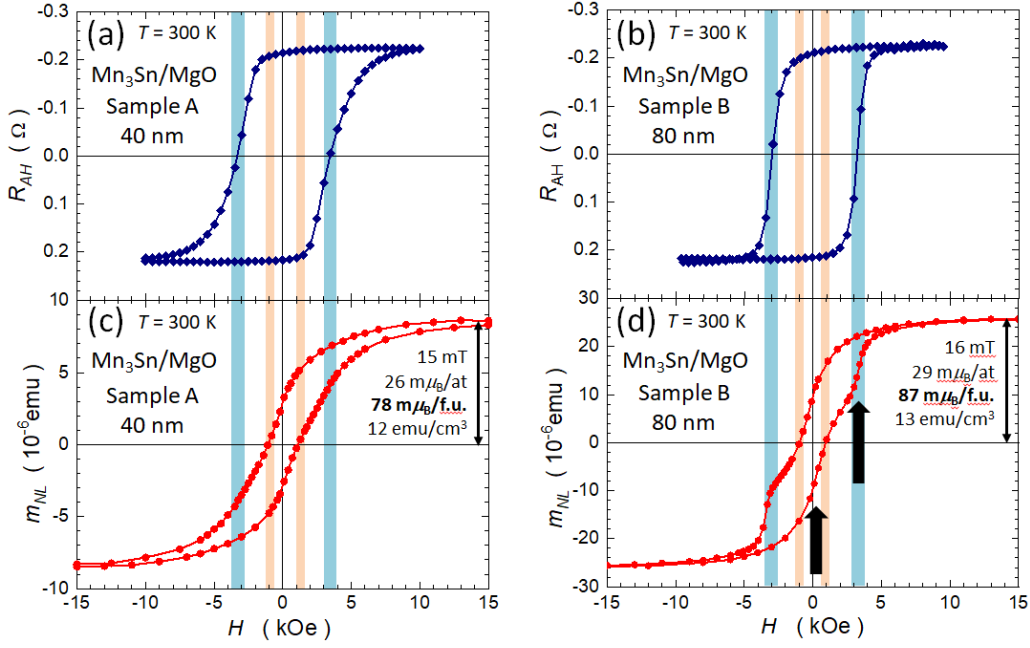

FIG 1. Comparison of the room temperature magnetic field, $H$, loops of the anomalous Hall resistance, $R_{AH}(H)$ (top panels), and of the nonlinear part of the magnetic loops, $m_{NL}(H)$ (bottom panels), for two representative $Mn_3Sn$/MgO samples. The $m_{NL}(H)$ are obtained from raw magnetometry results by subtracting a linear background term, $\alpha \cdot H$, individually adjusted for each curve to visually enforce saturation at high fields. Panels (a) and (c) correspond to Sample A ($Mn_{3.02}Sn_{0.98}$, 40 nm); panels (b) and (d) correspond to Sample B ($Mn_{3.09}Sn_{0.91}$, 80 nm). All measurements are performed with $H$ applied perpendicular to the sample plane (parallel to the kagome planes, $H \parallel [10\bar{1}0]$ of $Mn_3Sn$). The $m_{NL}(H)$ is shown in experimental units, corresponding to $5 \times 5$ mm$^2$ specimens. Vertical pale blue and orange bars indicate the coercive fields $H_C$, extracted from $R_{AH}(H)$ and $m_{NL}(H)$, respectively. The thick black arrows in panel (d) mark two distinct features in $m_{NL}(H)$, sometimes interpreted as 'double switching', although they originate from different contributions, as shown in the text. The magnetization values indicated by arrows in panels (c) and (d) represent apparent saturation levels obtained by attributing the entire magnetic moment to the $Mn_3Sn$ layer.

After such adjustment, the resemblance between $m_{NL}(H)$ and $R_{AH}(H)$ is only limited. The most notable discrepancies occur at weak fields region. For Sample A (panels a and c), $R_{AH}(H)$ reverses at approximately 3.3 kOe, while $m_{NL}(H)$ suggests a coercive field, $H_c$, of about 1.1 kOe – a threefold difference difficult to reconcile if both signals were to reflect the same magnetic switching event. For Sample B (panels b and d), the mismatch is even more pronounced. The $R_{AH}(H)$ loop exhibits a sharp, single switching around 3.1 kOe. In contrast, $m_{NL}(H)$ appears to show two distinct switching features: one that aligns with that of AHE at 3.3 kOe, and a second one near zero field (the $dm_{NL}(H)/dH$ derivative peaks at approximately ±310 Oe), a feature completely absent in the transport data. Since the $R_{AH}(H)$ response resembles that of well-characterized bulk $Mn_3Sn$ [20], the observed mismatch points strongly to unresolved parasitic contributions in the magnetometry signal.

In addition to altering the switching behavior and coercive fields, these parasitic components also distort the absolute values of magnetization extracted from such $m_{NL}(H)$. Enforced by the linear compensation apparent saturation values on the order of $10^{-5}$ emu may seem reasonable at first glance, but are implausibly large when considering the nanometer-scale thickness of the antiferromagnetic layers. Assuming the $m_{NL}(H)$ curves shown in Fig. 1 represent valid film data, the resulting saturation magnetizations exceed known bulk values [20] by at least a factor of 15. Such an outcome, if taken at face value, would imply a striking enhancement of magnetic properties – an observation that warrants careful consideration regarding its origin and underlying mechanisms.

Furthermore, the rounded shape of $m_{NL}(H)$ around zero field results in artificially low remnant magnetization values, $m_{REM}$, usually not exceeding one third of the saturation level. This, too, is inconsistent with $R_{AH}(H)$ data, which show a far better squarenes and near a full remanence.

The results considered in Fig. 1 demonstrate that the widespread practice of treating the substrate as an ideal diamagnet – approximated by a simple linear-in-field correction ($\alpha \cdot H$) adjusted to enforce saturation – can lead to significant errors in the determination of spontaneous magnetization, coercive fields, and magnetic susceptibility. This undermines the physical interpretation of magnetic data and risks masking genuine properties of the film under investigation. To address this issue, we propose and validate a practical framework for identifying and subtracting substrate-derived magnetic components on a per-sample basis, without requiring any prior substrate characterization. The need for such approaches is increasingly recognized in the literature [16,18,19,44,46] particularly in studies of AFM materials and other weakly magnetic systems. Our own analysis, focused on $Mn_3Sn$ films grown on MgO, reveals that apparent discrepancies between anomalous Hall and magnetization loops – such as those illustrated in Fig. 1 – are exclusively caused by incorrect data reduction, specifically by subtracting a strictly linear-in-field term that does not actually represent the true $m(H)$ dependence of the substrate. This underscores the importance of systematic artifact mitigation and motivates the detailed experimental characterization of MgO substrates presented below.

With this motivation, we now turn to a detailed experimental characterization of the magnetic signals originating from commercial MgO substrates.

**B. Magnetic properties of MgO substrates**

We begin this section by presenting the broader context of deviations from the expected ideal diamagnetic behaviour observed in commercially available MgO substrates, which are widely used for thin-film growth in studies of antiferromagnetic and topological materials such as $Mn_3Sn$. This issue is particularly critical, as MgO substrates, while offering undeniable structural advantages, indeed exhibit magnetic properties that deviate significantly from ideal diamagnetic behaviour – ultimately compromising both the quantitative and qualitative interpretation of magnetometry data.

In main panel of Fig. 2 we show the temperature dependence of the magnetic moment measured for a typical MgO substrate. The data reveal a strong temperature-dependent signal, with the total moment changing by more than $2\times 10^{-3}$ emu between 400 K and 2 K, reflecting its very paramagnetic origin - which directly contradicts an assumption of a temperature-independent diamagnetic character of MgO substrates. Notably, the magnetization switches sign – becoming strongly positive at low temperatures. The inset displays the central part of magnetic field isotherm $m(H)$ measured at 400 K for the same sample. Here, despite such elevated temperature, pronounced nonlinearity is evident within the ±5 kOe range, as highlighted by the overlaid straight line aligned to the linear part of $m(H)$ at magnetic fields above 10 kOe. Following the results from the main panel, the line represents the sum of the (dominating) diamagnetic response and the PM component of this specimen. These observations clearly demonstrate that MgO substrates cannot be regarded as ideal diamagnets over the temperature and field ranges typically relevant to thin-film magnetometry.

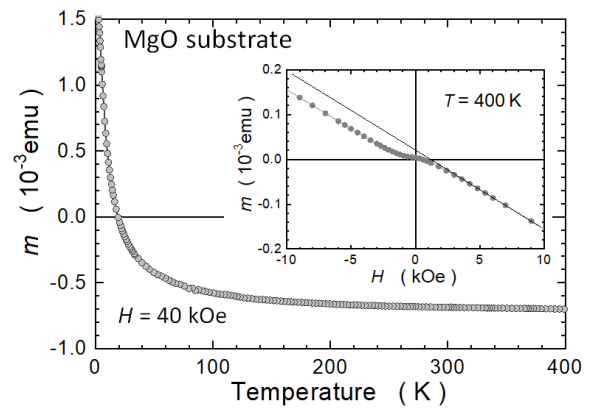

FIG. 2. Temperature $T$ dependence of the magnetic moment $m$ measured at magnetic field $H = 40$ kOe for an MgO $5 \times 5 \times 0.5$ mm$^3$ substrate, showing a strong temperature-dependent signal characteristic of a paramagnetic contribution. Inset: central part of the magnetic isotherm $m(H)$ measured at 400 K for the same sample. The straight line overlaid on the high-

field region illustrates the deviation from ideal linear behavior in low and intermediate fields.

In the following sections, we examine these deviations from the ideal diamagnetism in MgO substrates in greater detail, quantifying their impact relative to typical magnetic signals expected from thin films. We then propose and validate a set of non-destructive mitigation strategies, capable of eliminating as much as 97% of the substrate-related artifacts. This level of suppression - achieved without altering or damaging the sample - is shown sufficient to enable reliable and practically meaningful interpretation of magnetometry data in most experimental scenarios.

### B.1. The ferromagnetic-like contributions

We begin by analyzing the ferromagnetic-like (FML) contribution present in commercial MgO substrates. The main panel of Fig. 3 presents the nonlinear magnetic response, $m_{NL}(H)$, obtained for a series of nominally identical $5 \times 5 \times 0.5$ mm$^3$ MgO wafers sourced from the same vendor. For clarity, only one branch of each $m_{NL}(H)$ loop is shown; the central portion of the full hysteresis loop is illustrated in the top-left inset. To emphasize the deviation from the commonly assumed linear magnetic background, we intentionally apply a linear compensation of the form $\alpha \cdot H$ to all data. This highlights the extent to which the actual substrate response deviates from ideal diamagnetic behavior. As seen, each sample exhibits a distinct and substantial FML component, with saturation values, $m_{NL}^{sat}$, ranging from approximately 6 to 25 µemu, underscoring the considerable sample-to-sample variability observed even among substrates from the same batch. Such magnitudes of $m_{NL}^{sat}$ significantly exceed the magnetic responses typically expected from thin-film samples, especially antiferromagnetic films. To illustrate, these $m_{NL}^{sat}$ values translate into apparent magnetizations of up to a half of $\mu_B$/f.u. if interpreted as originating from a 20 nm Mn$_3$Sn film with the same footprint (as indicated on extra right-hand axis of Fig. 3).

The sample-to-sample variability demonstrates that no universal reference substrate can be reliably defined for straightforward background subtraction. Direct subtraction of an arbitrarily selected MgO reference curve risks introducing significant systematic errors. Furthermore, we highlight that the remnant magnetization, $m_{NL}^{REM}$, also varies considerably across samples, ranging here from 1.5 to 4 µemu. However, the ratio $m_{NL}^{REM}/m_{NL}^{sat}$, remains relatively constant, typically assuming a small value of 15–20%. Together with the rounded shape of $m_{NL}(H)$, this strongly suggests that FML is a non-uniform magnetic state without true long-range ferromagnetic order, composed of weakly interacting magnetic entities. Collectively, these regions behave as blocked superparamagnets (BSP), as shown below.

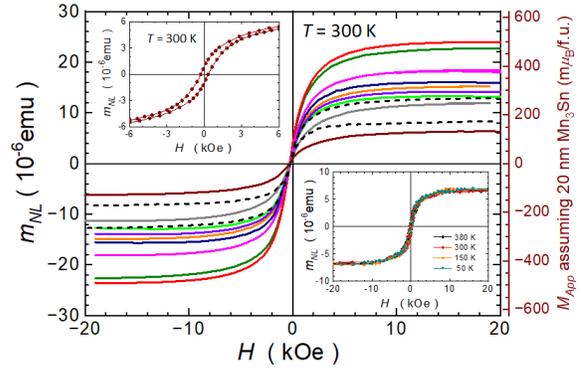

FIG. 3. Magnetic field $H$ dependence of the nonlinear component $m_{NL}(H)$ extracted from the experimental data of nominally identical MgO $5 \times 5 \times 0.5$ mm$^3$ substrates from a single manufacturing batch obtained by subtracting a linear background term, $\alpha \cdot H$, individually adjusted for each curve to visually enforce saturation at high fields. Solid/dashed lines in the main panel represent a set of (110)/(111)-oriented MgO substrates, respectively. For clarity, only one branch of each $m_{NL}(H)$ curve is shown. The additional right-hand Y axis shows the apparent magnetization $M_{App}$ one would infer if this signal were wrongly attributed solely to a 20 nm Mn$_3$Sn film of the same lateral dimensions. Top-left inset: central part of the complete $m_{NL}(H)$ hysteresis loop for a representative MgO substrate. Bottom-right inset: $m_{NL}(H)$ curves of an MgO substrate measured at various temperatures.

### B.1a Blocked superparamagnetism

The non-uniform magnetic nature of the FML component is confirmed by examining the temperature dependence of a representative MgO substrate in its remnant state, with other magnetic contributions effectively nullified. For this measurement, the specimen is first cooled in a magnetizing field down to 2 K. The field is then quenched to zero, bringing the sample to remanence. The thermoremnant moment ($m_{TR}$) is traced through successive temperature cycles according to the protocol shown in the inset of Fig. 4. Each subsequent cycle involves warming the sample to progressively higher temperatures followed by re-cooling back to 2 K - a refined variation of standard thermoremnant magnetization measurements designed to clarify the magnetic response [47,48].

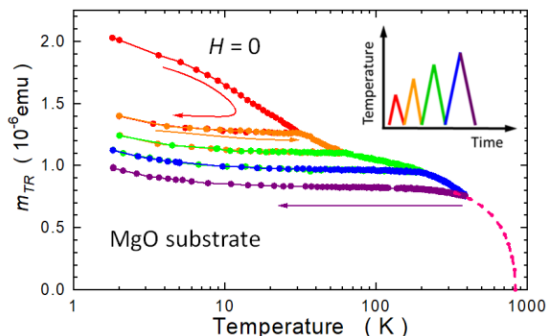

FIG. 4. Thermoremnant moment ($m_{TR}$) of MgO substrate during cyclic heating and cooling in zero magnetic field. The inset illustrates the applied temperature cycling procedure. The dashed curve represents the classical Langevin function $L(T)$, which describes the magnetization of non-interacting macrospins obtained in the classical limit ($J = \infty$) of the molecular field theory.

The resulting step-like behavior of $m_{TR}$, which remains essentially constant upon re-cooling (Fig. 4), provides clear evidence that $m_{NL}$ originates from blocked superparamagnetic entities. In contrast to uniform ferromagnets, where thermal demagnetization occurs smoothly [47], this discrete, ladder-like stepwise decrease pattern points to a systems containing blocked superparamagnetic clusters (macrospins) [49,50]. The dashed line in Fig. 4 represents the classical Langevin function $L(T)$, modelling the thermal evolution of magnetization for large magnetic moments. From this pseudo-fit, we estimate a characteristic coupling temperature for FML clusters of approximately 800 ± 100 K.

Notably, the $m_{TR}$ background remains stable over a broad temperature range (from 400 down to 10 K, as indicated by the last leg of $m_{TR}$ measurement), indicating that the FML component is an intrinsic property of the MgO substrate rather than a consequence of magnetic phase transitions. This weak temperature dependence – confirmed by direct magnetization measurements between 50 and 400 K (inset to Fig. 3) – proves highly advantageous in experimental practice. It enables reproducible compensation of the FML signal across the entire operational range of commercial magnetometry platforms, including Quantum Design's MPMS and PPMS systems, without the need for temperature-specific recalibration, and underpins our compensation strategy, as presented later in this section.

### B.1b Magnetic anisotropy

Another important observation is that the FML in MgO substrates exhibits significant magnetic anisotropy between in-plane and out-of-plane orientations. A typical example is presented in Fig 5. The data show that below about 10 kOe, the in-plane magnetization increases more steeply, indicative of easy-axis-like behavior, while the out-of-plane magnetization grows more gradually, corresponding to hard-axis-like behavior. The black symbols in Fig. 5 and the hatched area represent the difference between the two orientations, providing a direct measure of the anisotropy of the FML. This difference reaches a maximum around 1.5 kOe, with a magnitude of about 2 μemu here, or about 25% of the saturation value of the FML in other substrates. These findings reveal an additional complication introduced by the substrates themselves, which must be carefully considered in orientation-dependent analyses of thin-film magnetization data.

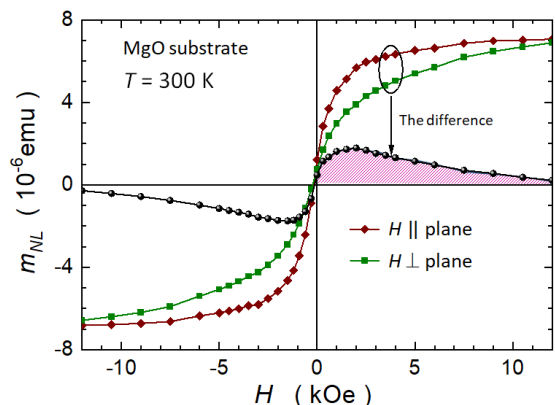

FIG. 5 Nonlinear magnetic moment $m_{NL}(H)$ of a $5 \times 5$ mm$^2$ MgO substrate measured at 300 K in two orthogonal field orientations: in-plane (brown diamonds) and out-of-plane (green squares). The black symbols represent the difference between the two curves, providing a direct measure of the magnitude and field dependence of the magnetic anisotropy of the ferromagnetic-like component (FML) in the substrate.

### B.1c Mitigation strategy: Internal Reference Method (IRM)

We begin by noting that the analysis presented in Appendix B reveals that the FML contribution observed in MgO substrates originates from localized sources at or near the epi-ready surface and exhibits little correlation with substrate thickness or bulk volume. While this identification is important from a fundamental standpoint, it offers limited practical benefit for thin-film applications. Surface etching cannot be applied in this context, as exposure to aggressive chemicals irreversibly damages the pristine epi-ready surface, rendering the substrate unsuitable for epitaxial growth.

On the other hand, a powerful experimental method for eliminating the substrate contribution directly at the measurement stage - based on *in situ* magnetic compensation using a matching compensating material - has already been developed by the present authors [40]. Unfortunately, this approach is severely hindered in the case of MgO due to the pronounced sample-to-sample variability in its magnetic response, which precludes reliable matching of the compensating elements to the substrate of a given specimen. This is a significant limitation, as the *in situ* method has proven remarkably effective in achieving high-precision magnetometry of weak magnetic signals in thin layers and dilute magnetic semiconductors [41] and, most importantly in the present context, antiferromagnetic thin films [42,51,52].

We therefore put forward an alternative strategy that can address the unique challenges posed by MgO substrates. Fortunately, the relatively weak temperature dependence of the FML component offers a viable path forward. This forms the basis of the internal reference method (IRM), which enables the removal of the FML component by exploiting the fact that the magnetic signal from the film usually vanishes at elevated temperatures, while the substrate contribution remains largely unchanged. Specifically, for each individual sample, a high-temperature reference loop $m_R(H)$ is recorded a reference temperature, $T_R$- say at 400 K for $Mn_3Sn$/MgO structures - and then subtracted from the lower-temperature measurement. This procedure effectively isolates the temperature-dependent magnetic response of the thin film while eliminating parasitic substrate signals.

Crucially, the reference and measurement loops must be acquired under identical experimental conditions, including field orientation, mounting configuration, and instrument geometry. This strict consistency is essential, as absolute values in SQUID magnetometry are sensitive to mechanical alignment, and the magnitude and shape of the FML vary significantly between substrates – even within the same manufacturing batch (see Fig. 3). Consequently, no universal background or pre-characterized reference curve can be applied: a dedicated $m_R(H)$ measurement is required for each sample to achieve meaningful correction.

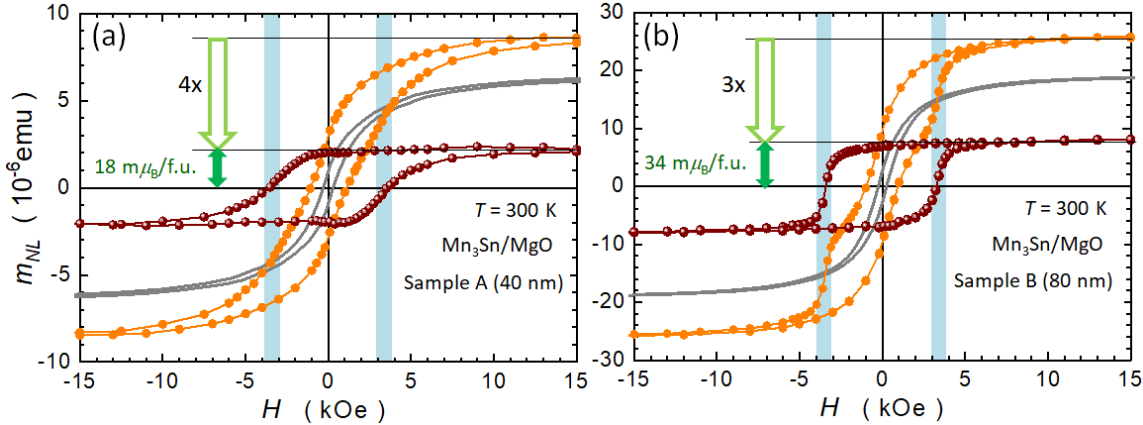

FIG. 6 (Brown bullets) Non-linear parts of the magnetization loops, $m_{NL}(H)$, after application of the internal reference method (IRM), $m_{IRM}(H)$, for the same two $Mn_3Sn$ layers shown in Fig 1. For each sample, the correction is performed individually by subtracting its high-temperature reference loop, $m_R(H)$, measured at 400 K (grey lines), thereby eliminating the sample-specific ferromagnetic-like component (FML) originating from the MgO substrate. The vertical pale blue bars, as in Fig. 1, indicate the coercive fields determined from anomalous Hall effect (AHE) measurements. The IRM-corrected $m_{IRM}(H)$ loops exhibit significantly improved agreement with the AHE data (not shown here; see Fig. 1), both in coercive fields and loop shape. The thick green arrows indicate the reduction of the apparent magnetic saturation, which after IRM processing now represents the true saturation level of the $Mn_3Sn$ layers. In this figure, all $m_{IRM}(H)$ curves have been artificially levelled at high fields (by adjusting linear compensation factors α) to clearly emphasize the magnitude of changes recovered in the intrinsic field dependence of the films and to facilitate visual comparison with the AHE results. This adjustment does not reflect the actual high-field behavior, as proper correction of the paramagnetic contribution from the MgO substrate will be addressed in the following section.

The effectiveness of the IRM procedure is demonstrated in Fig. 6, which presents the corrected results (brown bullets) for the same two samples shown earlier in Fig. 1. It is immediately evident that the presence of the FML component (grey lines) is responsible for the unphysically vertically stretched magnetometry data (orange points) and/or the appearance of additional magnetic "events" [such as the "double switching" observed in panel (b)], which arise under the assumption of ideal diamagnetic behavior of the MgO substrate.

Elimination of this spurious FML component restores the qualitative consistency between the magnetometry and magnetotransport results, as established for bulk $Mn_3Sn$. In particular, the coercive field values obtained after IRM correction are significantly larger and now coincide with those given by AHE measurements – thus indicating that the commonly invoked explanations for discrepancies between AHE and $m(H)$ measurements in thin films may warrant closer examination.

Finally, it is worth noting the clear difference in $m_R(H)$ values between the two samples shown in Figs. 1 and 6, which evidences the necessity of determining an individual high-temperature reference $m_R(H)$ curve for each specimen. Only through such sample-specific correction can the intrinsic $m(H)$ response of the film be reliably extracted from magnetometry

Another important outcome of eliminating the FML contribution is the substantial improvement in the agreement between the absolute values of the spontaneous magnetization obtained for the $Mn_3Sn$ thin films and those known from bulk magnetometry. As shown in Fig. 6, application of the IRM procedure (indicated by thick arrows) reduces the apparent saturation magnetization by a factor of four for the Sample A and by a factor of three for the Sample B. After the correction, the extracted magnetization values fall within the expected range for thin layers of $Mn_3Sn$, yielding approximately 18 m$\mu_B$/f.u. (about 2.6 mT) for Sample A and 34 m$\mu_B$/f.u. (about 5.1 mT) for Sample B [24].

Importantly, the IRM procedure also significantly reduces the paramagnetic contribution, with its effectiveness increasing as the measurement temperature approaches that of the high-temperature reference. This approach works particularly well when the $m_{PM}(H)$ behavior remains approximately linear. However, obtaining physically meaningful magnetization curves over a broader temperature range - especially at temperatures substantially lower than the IRM reference - and across the full magnetic field range requires explicit treatment of the second parasitic contribution from the MgO substrate: the paramagnetic component.

**B.2 Mitigation strategy of paramagnetic component**

The presence of a pronounced paramagnetic component, $m_{PM}$, in MgO substrates (as well as in other oxide crystals in a lesser degree) is associated with the presence of transition metal (TM) ions – primarily Fe, but also Mn, Cr, V, and Ca – distributed throughout the bulk of the material [19]. At room temperature (~300 K), this contribution may appear relatively weak, especially when compared to the dominant diamagnetic response of the MgO lattice. Nevertheless, it constitutes a significant source of interference when attempting to determine trends or absolute magnetization values in high fields for thin magnetic layers.

Because $m_{PM}$ originates from a mixed population of TM ions [19], it cannot be reliably described by a simple Curie–Weiss law. To complicate matters further, and similarly to the FML, the magnitude of $m_{PM}$ varies considerably from substrate to substrate, making it impossible to define a universal reference for eliminating this PM background. Consequently, there is once again a need for a dedicated, non-destructive method to estimate the substrate-related PM contribution after deposition – specifically for samples that already contain magnetic films and are considered valuable for further investigation.

As in the case of the FML correction, this approach requires some additional measurements. However, in this instance they must be performed at very low temperatures. The core idea and practical utility of the method proposed here – hereafter referred to as the Differential Isotherm Method (DIM) – are based on a key observation: for most materials that contain two or more qualitatively distinct magnetic components, the paramagnetic contribution is typically the only one that changes significantly with temperature in the low-temperature (e.g., liquid helium) regime. If the temperature dependence of other magnetic contributions (e.g., diamagnetic, FM, or BSP) can be neglected in comparison to the PM component, then the difference between two magnetic isotherms, $m(H)$, measured at two nearby low temperatures, effectively isolates the change in the paramagnetic signal [53]. This differential approach has been successfully used to separate $m_{PM}$ from BSP or FM signals in a variety of systems, including DMS films [54], bulk crystals [55], and nanowires [49,56].

To demonstrate both the applicability and limitations of the DIM, we analyze the *m*(*H*) curves of the as-supplied and subsequently HCl-etched MgO substrates (as introduced in the previous section), measured at 2 K and 5 K - i.e., in the temperature range where the paramagnetic response is particularly pronounced. The results are presented in Fig. 7. These results again demonstrate the substantial strength of the bulk-related PM contribution in MgO substrates, as the *m*(*H*) curves obtained for the as-supplied sample (open symbols) and for the etched sample are practically identical. Far more important for the DIM, however, is the fact that despite the minor differences observed between the full *m*(*H*) data before and after etching, the differences between the measured isotherms at 2 K and 5 K – denoted as $\Delta M_{exp}$ (gray and pink bullets, respectively) – are practically indistinguishable. This confirms that $m_{PM}$ originates from TM impurities distributed throughout the bulk, rather than from surface contamination. These differences are highlighted graphically in Fig. 7 using the hatched area and are attributed solely to the PM contribution. As shown below, modelling this $\Delta M_{exp}$ allows us to extract quantitative parameters characterizing $m_{PM}$.

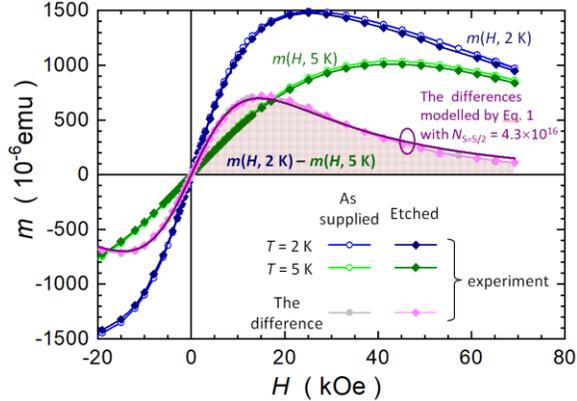

FIG. 7 Magnetic isotherms *m*(*H*) measured at 2 K (blue) and 5 K (green) for an MgO 5 × 5 mm² substrate, comparing as-supplied (open symbols) and HCl-etched (full symbols) results. The difference between the 2 K and 5 K curves, $\Delta M_{exp}(H)$ (gray and magenta points, respectively), is identical for both cases, confirming that the paramagnetic contribution originates from bulk TM impurities. The purple solid line shows the differential magnetization $\Delta M_{theo}$, calculated within the Differential Isotherm Method (DIM) as the difference between two $S = 5/2$ Brillouin functions, and fitted to the experimental $\Delta M_{exp}(H)$. The extracted spin concentration $N_{S=5/2} = 4.3 \times 10^{16}$ is consistent with an Fe concentration of approximately 70 ppm provided by the vendor of the investigated substrate.

First, and somewhat unexpectedly, we find that despite its complex microscopic origin, the experimental $\Delta M_{exp}$ observed in MgO substrates can be accurately modelled using the classical Brillouin function, $B_S(T, H)$, assuming that the signal arises entirely from a single species of paramagnetic ions with orbital momentum $L = 0$, i.e., total angular momentum $J = S$. Under this assumption, the differential magnetization is given by:

$$\Delta M_{theo} = g\mu_B N_S S \Delta B_S(T,H) = g\mu_B N_S S \cdot [B_S(2\,\mathrm{K}, H) - B_S(5\,\mathrm{K}, H)], \quad (1)$$

where *g* is the Landé *g*-factor, *S* is the spin quantum number, and $N_S$ is the number of contributing spins. $B_S$ assumes the form:

$$B_S(x) = \frac{2S+1}{2S} \coth\left(\frac{2S+1}{2S}x\right) - \frac{1}{2S}\coth\left(\frac{1}{2S}x\right), \quad (2)$$

with *H* and *T* tied in *x* by:

$$x = S\frac{g\mu_B H}{k_B T}, \quad (3)$$

($k_B$ is the Boltzmann constant). In the simplest implementation of the DIM, the spin value *S* is assumed, leaving $N_S$ as the only adjustable parameter, which effectively acts as a scaling factor – thus avoiding the need for nonlinear fitting. Remarkably, adopting $S = 5/2$ yields excellent agreement between the theoretical $\Delta M_{theo}$ and the experimental $\Delta M_{exp}(H)$, as shown in Fig. 7 by the purple line, which is calculated for $N_{S=5/2} = 4.3 \times 10^{16}$. This value is generally consistent with the Fe concentration specified by the manufacturer (70 ppm), though some variability is observed, with $N_{S=5/2}$ values ranging between $4.2 \times 10^{16}$ and $4.7 \times 10^{16}$ spins across the range of investigated samples.

We find that, owing to the substantial magnitude and strong temperature dependence of $m_{PM}$ at low temperatures, the DIM can reliably quantify this signal – even when applied directly to the final heterostructure, i.e., after the deposition of the magnetic layer. Naturally, this raises the question of how well such a parametrization reflects the true magnitude of $m_{PM}$ at elevated temperatures (e.g., in the 200–400 K range), which is most relevant for current studies of AFM and AM systems.

To address this issue, we analyze the *m*(*T*) curve measured at elevated temperatures for the same HCl-etched bare MgO substrate, thus eliminating the influence of any FML component. As shown in Fig. 8, even with the approximate parametrization derived from low-temperature DIM analysis, it is possible to suppress approximately 86% of the substrate's paramagnetic component at high temperatures. While this figure might appear modest in isolation, it must be emphasized that the IRM correction also removes a substantial fraction of $m_{PM}$ – especially when the measurement temperature is close to the IRM reference temperature.

To fully leverage the complementary strengths of the IRM and DIM approaches in removing $m_{PM}$, we implement a two-step correction scheme as described below. It is illustrated in Fig. 8, however for illustrative reasons the $T$-dependence of $m$ is considered:

(i) In the first step, the paramagnetic component $m_{DIM}(H, T)$ is estimated using the DIM parametrization established from low-temperature measurements (as shown in Fig. 7). (These values are marked by magenta arrows in Fig. 8). The $m_{DIM}(H, T)$ corrections are then subtracted from the raw data: $m_{(i)}(H, T) = m(H, T) - m_{DIM}(H, T)$, both at $T_R$ and at all other measured temperatures.

(ii) In the second step, the IRM procedure is applied: the reference curve $m_{(i)}(H, T_R)$ – already corrected for the PM contribution in step (i) - is subtracted from each DIM-corrected $m_{(i)}(H, T)$. This removes the temperature-independent FML component and, as a bonus, also the part of $m_{PM}(H, T_R)$ not corrected in step (i) - marked in Fig. 8 by a yellow hatched area - thus substantially improving the effectiveness of $m_{PM}$ removal in step (i).

As illustrated in Fig. 8, this combined correction strategy suppresses the residual paramagnetic signal near room temperature to just 2–3% of its original magnitude - i.e. from $m_{PM}(300 \text{ K}, 40 \text{ kOe}) = 50$ μemu to $\Delta m_{PM} \cong 1.5$ μemu, a reduction by more than a factor of 30. As a result, the fidelity of the reconstructed $m(H)$ specific to the investigated films at elevated temperatures and high magnetic fields improves by approximately a factor of 50.

Notably, the general effectiveness of DIM stems from its straightforward physical basis. While in this work we have illustrated the method assuming a single $S = 5/2$ spin species, the formalism can readily be extended to incorporate other spin states or multiple spin species simultaneously [55]. An alternative approach involves treating the spin quantum number $S$ as an adjustable fitting parameter (i.e. an effective spin $S_{\text{eff}}$), depending on the desired precision and modeling assumptions. If needed, manual tuning of parameters – such as increasing $N_S$ by ~10% - can improve compensation in specific temperature windows (e.g. above ~150 K). Naturally, such adjustments should be made with caution and supported by appropriate validation tests.

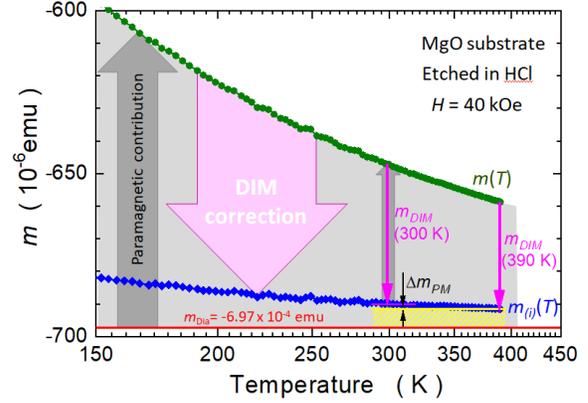

FIG. 8 Quantitative illustration of the effectiveness of the combined Internal Reference Method (IRM) and Differential Isotherm Method (DIM) correction in suppressing the paramagnetic (PM) contribution originating from the bulk of an MgO substrate. Green circles: temperature dependence of the magnetic moment $m(T)$ measured for a $5 \times 5$ mm$^2$ HCl-etched MgO substrate at $H = 40$ kOe. Red solid line: net diamagnetic moment $m_{\text{Dia}} = -697$ μemu, established independently (see Appendix C), which serves as the reference level for quantifying the PM component $m_{PM}(T)$. Its magnitudes are indicated by grey arrows and the shaded area. Blue diamonds: DIM-corrected data $m_{(i)}(T) = m(T) - m_{DIM}(T)$. $m_{DIM}(T)$ is calculated using the Brillouin function (Eq. 2) with $S = 5/2$ and $N_S = 4.3 \times 10^{16}$ in this case. Black distance arrows indicate the remaining uncorrected PM moment $\Delta m_{PM} \cong 1.5$ μemu at $T = 300$ K: $m_{(ii)}(300 \text{ K}) = m_{(i)}(300 \text{ K}) - m_{(i)}(390 \text{ K})$. When compared to the total PM contribution at 300 K, $m_{PM}(300 \text{ K}) \cong 50$ μemu, it yields an effectiveness in the reduction of the PM component of at least 97%.

We emphasize that the DIM alone offers a practical and robust framework for quantitatively evaluating the paramagnetic contribution originating from the bulk of MgO and similar oxide substrates. Importantly, under its specific assumptions, the method relies solely on low-temperature magnetization data and requires no prior magnetic characterization of the substrate before deposition. As with the IRM approach, this makes DIM particularly valuable for retrospective application – especially when subsequent measurements indicate the need for more precise magnetic analysis.

However, it should also be noted that DIM, by design, will identify and quantify any purely PM contribution present in the measured sample - including paramagnetic signals originating from the film itself. On one hand, this feature adds to the method's universality and sensitivity in detecting weak PM components. On the other hand, when applying DIM to remove such signals from experimental data, particular care is required to ensure that meaningful PM contributions intrinsic to the sample are not inadvertently subtracted. In this context, the method should not be applied blindly but with careful judgment and awareness of the underlying physical properties of the system under investigation.

### C. Evaluation of the whole compensation scheme

We now present the final results obtained by applying the whole correction scheme to the magnetometry results of the two $Mn_3Sn$/MgO thin films introduced at the beginning of this study. These results, already converted to the magnetization $M$ are shown in Fig. 9. As seen, the corrected $M(H)$ loops exhibit excellent agreement with the switching fields determined from AHE measurements: both the coercive fields and the squareness of the $M(H)$ loops now match the transport data. The resulting saturation magnetization values are approximately two to three times larger than those reported for bulk crystals, which is consistent with previous studies highlighting the influence of epitaxial strain on the magnetic properties of $Mn_3Sn$ [24,57].

Interestingly, the slope of the $M(H)$ curves in the high-field regime is essentially identical for both samples, despite their differing thickness (40 nm vs 80 nm) and Mn content. This behavior is expected, as the Zeeman effect should be thickness-independent in first approximation, so the spin canting should be identical when considered in absolute magnetization units. Notably, the extracted high-field susceptibility, $\chi_{Mn_3Sn}$ = 11 ± 1 $\mu_B$/f.u./10 kOe for these samples, closely matches the value reported for bulk samples (12.5 ± 0.5 $\mu_B$/f.u./T; c.f. Fig. 2(d) in Ref. [57]), further validating the accuracy and applicability of our proposed correction scheme.

In this context, it is important to note that applying IRM alone is insufficient to recover physically meaningful high-field behavior. While IRM yields reliable estimates of the spontaneous magnetization, only by additionally applying DIM - illustrated by the curved arrows in Fig. 9 - can one restore full agreement with reference data in the high-field limit. This final step of our procedure significantly enhances the fidelity of magnetic characterization in strained $Mn_3Sn$ thin films and underscores the utility of combining IRM and DIM in tandem. These findings form the basis for the general conclusions summarized below.

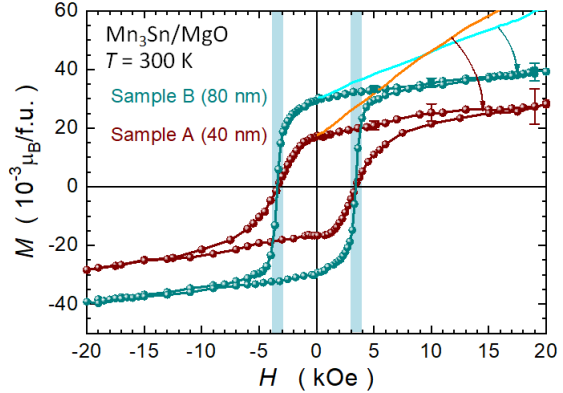

FIG. 9 Magnetization as a function of magnetic field $H$, obtained after applying both the Internal Reference Method (IRM) and the Differential Isotherm Method (DIM) to correct for non-diamagnetic background contributions from commercial MgO substrates, for the two exemplary $Mn_3Sn$ thin films investigated in this study. Vertical pale blue bars indicate the coercive fields determined from anomalous Hall effect measurements, as in Fig. 1. Thick solid lines in the first quadrant represent data corrected using IRM alone. The curved arrows illustrate the additional correction introduced by DIM, highlighting the improvement in high-field agreement. The error bars indicate the uncertainty in determining the exact magnitude of the paramagnetic component present in MgO, as shown in Fig. 8.

### IV. CONCLUSIONS

This study addresses a key challenge in thin-film magnetometry: the mitigation of significant, magnetic artifacts originating from commercial oxide substrates, taking MgO as the sole example. The ferromagnetic-like component is attributed to blocked superparamagnetic entities residing near or at the substrate epi-ready surface, while the paramagnetic response stems from dilute bulk impurities. Using then epitaxial $Mn_3Sn$/MgO films as a model system we show that standard linear background subtraction fails to recover intrinsic magnetic properties. We therefore introduce and validate two complementary, non-destructive, correction procedures that do not require pre-characterization of the substrate: the Internal Reference Method (IRM), which suppresses a weak-field ferromagnetic-like component, and the Differential Isotherm Method (DIM), which quantifies and removes the paramagnetic background. Applied in tandem, these methods substantially improve consistency between magnetometry and transport data and enable reliable extraction of coercive fields, spontaneous magnetization and absolute values of high field magnetic susceptibility (spin canting).

We also demonstrate that frequently reported discrepancies between anomalous Hall and magnetization loops can arise just from these substrate contributions, not from domain-related-alike effects in these films. While our work focuses on Mn$_3$Sn on MgO, the method is broadly applicable to antiferromagnets, altermagnets, 2D materials and many other systems stabilized on crystalline semiconductor substrates. By systematically addressing substrate-induced distortions, this work serves both as a practical guide for high-fidelity magnetometry and as a cautionary reminder that structurally perfect substrates may still produce magnetic signals that obscure or mimic intrinsic film.

## ACKNOWLEDGMENTS

The authors acknowledge Tomohiro Uchimura, Shun Wakabayashi, Ju-Young Yoon, Yutaro Takeuchi, and Shunsuke Fukami at Tohoku University for providing research materials and for helpful discussions. This work was partly supported by TUMUG Support Program from Center for Diversity, Equity, and Inclusion, Tohoku University.

## APPENDIX A: SUBSTRATE AND HANDLING ARTIFACTS IN MAGNETOMETRY

The main body of this work focuses on MgO substrates, which are among the most commonly used crystalline oxides in thin-film research. However, MgO is by no means unique in producing non-negligible background signals. Other widely used substrates - including LSAT, MgAl$_2$O$_4$, SrTiO$_3$, and Y:ZrO$_2$ - are also known to exhibit ferromagnetic-like or paramagnetic behavior, often contradicting the assumption of purely diamagnetic response It manifests as a sigmoidal hysteresis loop that saturates around 1 T, with a rather small coercive field typically between 15 and 25 mT. On average, its overall magnitude reaches approximately $10^{-5}$ emu, posing additional challenge for accurate interpretation of the thin-films intrinsic magnetism [19,58].

Beyond the substrate itself, contamination and handling artifacts can introduce similar or larger spurious signals [12,13,58]. Residual ferromagnetic particles introduced during cutting, grinding, or polishing, or transferred from metal tools (particularly ferrous ones), can add broad hysteresis loops with coercive fields from 10 mT to 100 mT and remnant magnetization reaching 10–20% of the saturation value [58], sometimes exceeding $10^{-4}$ emu at saturation [13]. Even the use of deformed or worn out plastic drinking straws as sample holders can produce false ferromagnetic-like signals exceeding $10^{-6}$ emu [13]. Particularly dangerous is an inhomogeneous distributions of these contaminants – it builds up the spurious, not-really-existing response and may cause additional complications – such as imperfect sample alignment.

Other artifacts are caused by residual flux in the superconducting coil and background signals from mounting materials. They become largely problematic at low fields [43]. Both can further affect measured coercive fields and displace hysteresis loops [14,43]. Residual flux in superconducting magnets and magnetic background from sample mounting materials can further contribute to spurious low-field signals [56,58]. These effects are especially problematic in studies involving weakly magnetic or compensated systems, where even minute background distortions may be misinterpreted as intrinsic properties.

To highlight the variety and magnitude of known spurious signals, we summarize several representative values reported in the literature in Table 1 below.

TABLE I. Measured signal strength of artifacts in magnetometry

| Reference | Artifact Type | Measured Signal (emu) | Characteristics |
| --- | --- | --- | --- |
| [13] | Magnetic contamination from tweezers | up to $10^{-4}$ | Magnetic signal observed on substrates contaminated by stainless steel tweezers, varies by contamination level. |
| [15] | Magnetic contamination from cleaving - Diamond on stainless steel rod (SS stencil) | up to $4\times10^{-6}$ | Ferromagnetic residues introduced by diamond-tipped cleaving tools mounted on stainless steel rods, dependent on material and handling conditions. |
| [15] | Magnetic contamination from cleaving - SS blade | up to $1.4\times10^{-5}$ | Metallic debris from stainless steel blades can adhere to the substrate, introducing extrinsic signals. |
| [13] | Kapton® tape residue | $\approx 5\text{-}15\times10^{-6}$ | Ferromagnetic-like contributions observed in Kapton® tape, varying with temperature and tape handling. |
| [13] | Drinking straw deformation | $\approx 10^{-6}$ | Mechanical deformation of straws used in sample mounting can induce weak magnetic artifacts. |
| [56] | Adhesive used for substrate mounting in MBE | $\approx 10^{-6}$ | Residual magnetic signal from adhesives used in MBE processing, varies with type and thickness. |

## APPENDIX B: EXPERIMENTAL VERIFICATION OF FML ORIGIN IN MgO SUBSTRATES

This appendix presents two complementary experiments designed to verify the origin of the ferromagnetic-like (FML) contribution observed in MgO substrates commonly used for thin-film growth. Combined, both experiments explicitly demonstrate that FML originates solely from the epi-ready surface of the MgO substrates.

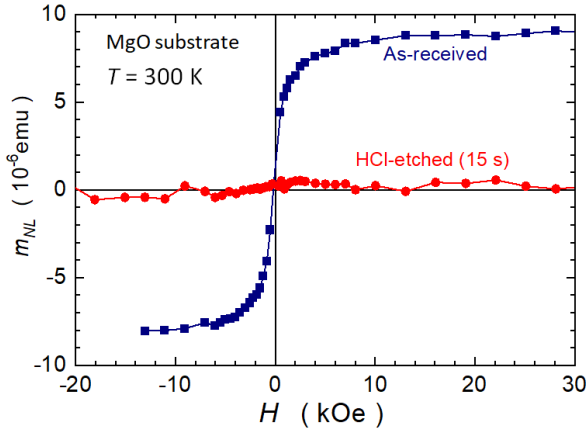

FIG. 10 Comparison of the non-linear parts, $m_{NL}$, of the magnetic curves measured for a $5 \times 5$ mm² MgO substrate at 300 K before (navy squares) and after brief, 15 s etching in HCl (red circles). A linear compensation, $a \times H$, is applied to both $m(H)$ curves to flatten their high-field dependence, thereby emphasizing the effect of etching on the magnetic response.

In the first experiment (Fig. 10), magnetic hysteresis loops are measured for a pristine $5 \times 5$ mm² MgO substrate (navy squares) and after briefly etching the substrate surface in hydrochloric acid (HCl) for 15 seconds (red circles). To clearly see the change, a linear background proportional to $H$ is subtracted from both datasets to bring up non-linear contributions clearly. As seen in the figure, the original FML component present in the as-delivered substrate is entirely eliminated after surface etching, conclusively identifying the substrates' surface as the exclusive source of this component.

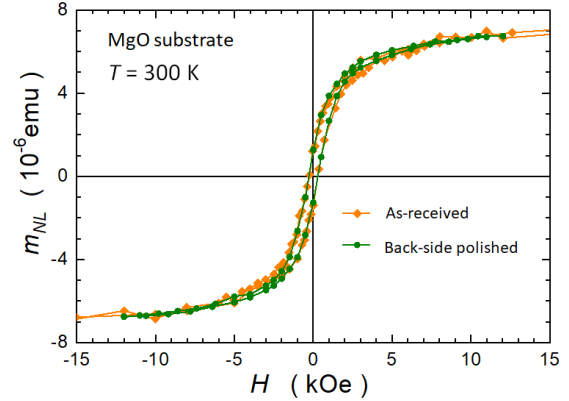

FIG. 11 Comparison of the nonlinear components $m_{NL}(H)$ of the magnetic response measured at 300 K for a $5 \times 5$ mm² MgO substrate before (orange diamonds) and after mechanical polishing of the non-epi-ready backside (green circles). Linear background corrections of the form $\alpha \cdot H$ are applied to the raw data to isolate and highlight the nonlinear component of the magnetic response. The unchanged FML after polishing demonstrates that this contribution originates exclusively from the intact epi-ready surface.

In a follow-up experiment (Fig. 11), we test whether the FML originates from both major surfaces of the MgO substrates, and whether mechanical polishing of the non-epi-ready (backside) surface could reduce this spurious contribution. Magnetic loops measured at room temperature before and after mechanical polishing of the backside of another MgO substrate are compared. The mechanical polishing removes approximately 20% of the substrate mass while intentionally leaving the epi-ready surface intact. The results clearly show that the backside polishing polishing does not reduce or alter the FML component, demonstrating that this contribution originates exclusively from the epi-ready side.

It is noteworthy that despite clearly identifying the epi-ready surface as the origin of the FML component, chemical etching cannot be recommended as a practical method for its removal. Any chemical exposure irreversibly alters the pristine surface condition, rendering substrates unsuitable for epitaxial thin-film growth. Thus, these experiments primarily serve diagnostic purposes rather than providing a practical substrate-cleaning methodology.

**APPENDIX C: DETERMINATION OF THE INTRINSIC DIAMAGNETIC SUSCEPTIBILITY OF MgO SUBSTRATES**

Determining the pure diamagnetic magnetic moment of MgO substrates is not straightforward due to the presence of two additional magnetic contributions: (i) a ferromagnetic-like component originating from the epi-ready surface of the substrate, and (ii) a paramagnetic (PM) response arising from transition metal impurities distributed throughout the substrate volume.

To isolate the diamagnetic contribution, we use the same substrate in which the FML component has been removed by chemical etching, as shown in Fig. 10 in Appendix B. To eliminate the PM contribution to the diamagnetism of the MgO substrate, we take advantage of the PM's characteristic temperature dependence, which follows approximately $1/T$ behavior and thus vanishes in the infinite-temperature limit. By performing $T$-dependent measurements at high-field and plotting $m(T)$ versus $1/T$, the extrapolated value of the magnetic moment at $1/T \to 0$ yields the diamagnetic moment (and ultimately the diamagnetic magnetization) of the substrate. The result of this procedure is presented in Fig. 12.

The determined diamagnetic susceptibility, $\chi_\mathrm{MgO} = -4.0 \times 10^{-7}$ emu/g/Oe, includes the correction factor $\gamma$, which accounts for the MPMS magnetometer's sensitivity to sample shape [59]. This measurement is performed in the parallel configuration, for which $\gamma = 0.983$ [14]. The practical value of $\chi_\mathrm{MgO}$ for $5 \times 5$ mm² substrate plates (not accounting for the $\gamma$ factor) is $-3.98 \times 10^{-7}$ emu/g/Oe and $-4.12 \times 10^{-7}$ emu/g/Oe for the in-plane and perpendicular configurations (for which $\gamma = 1.03$), respectively.

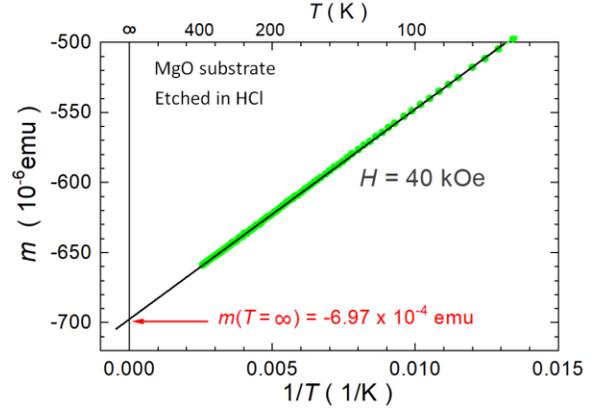

Fig. 12 High-field magnetic moment of the HCl-etched MgO substrate ($5 \times 5 \times 0.5$ mm³, mass 44.4 mg) plotted as a function of inverse temperature, $T$. The measurement is performed in an applied magnetic field of $H = 40$ kOe. Linear extrapolation to $1/T \to 0$ yields an intercept $m(T = \infty) = -6.97 \times 10^{-4}$ emu, corresponding to the diamagnetic moment of the substrate. This allows calculation of the diamagnetic susceptibility $\chi_\mathrm{MgO} = -4.0 \times 10^{-7}$ emu/g·Oe.

.